\documentclass[aip,jcp,reprint,notitlepage,amsmath,amsfonts,amssymb,floatfix,superscriptaddress]{revtex4-2}
\usepackage{physics}
\usepackage{graphicx}
\graphicspath{{figures/}}
\usepackage{dcolumn}
\usepackage{bm}
\usepackage[pdfauthor={Aditya Dev},
            pdftitle={Neutrino Mass Models},
            pdfkeywords={Neutrino mass, Dirac particle, Majorana Particle, CPT, Mass models, Neutrino Oscillations}
            ]{hyperref} 
\usepackage{xcolor}
\usepackage{amsfonts}
\definecolor{color1}{RGB}{0,0,90} 
\definecolor{color2}{RGB}{0,20,20} 
\hypersetup{
	hidelinks,
	colorlinks,
	breaklinks=true,
	urlcolor=color2,
	citecolor=color1,
	linkcolor=color1,
	bookmarksopen=false,
}

\begin{document}

\title[]{Neutrino Oscillations and Mass Models}

\author{Aditya Dev}
 \email{adityadev.iisermohali@gmail.com}
\affiliation{Department of Physical Sciences, Indian Institute of Science Education and Research Mohali}%

\date{\today}

\begin{abstract}
These notes provide a  review of various neutrino mass model and their implications for particle physics and the Standard Model.
 We discuss how mass terms are incorporated into the Standard Model, including the Dirac mass term and the Majorana mechanism. 
 We explore experimental evidence supporting the existence of non-zero neutrino mass and develop the formalism of neutrino oscillations. 
 Additionally, we provide a brief overview of some famous radiative neutrino mass models.  This notes draws heavily from the references given below and can serve as a valuable resource for students and new researchers interested in the field of neutrino physics.
\end{abstract}

\maketitle
\tableofcontents
\vspace{20pt}

\section{\label{sec:intro}Introduction}

Neutrinos are elementary particles that are electrically neutral and interact only weakly with matter. They were first postulated by Wolfgang Pauli in 1930 to explain the apparent violation of energy conservation in certain types of radioactive decay. Since then, neutrinos have been detected in a variety of astrophysical and laboratory settings, including nuclear reactors, particle accelerators, and the cosmic microwave background.

For many years, it was assumed that neutrinos were massless, due to their weak interactions and the absence of any direct evidence for their mass. However, in the late 1990s, a series of experiments provided the first evidence that neutrinos have non-zero mass. This discovery was a major breakthrough in particle physics and has led to a renewed interest in understanding the properties of neutrinos.

There are several models of neutrino mass that have been proposed over the years, each with its own strengths and weaknesses. These models include the seesaw mechanism, which postulates the existence of heavy right-handed neutrinos, and the radiative seesaw mechanism, which explains the smallness of neutrino mass through radiative corrections. Other models include the inverse seesaw mechanism, the low-scale leptogenesis scenario, and the Majorana model.

Neutrinos have been demonstrated to be priceless probes when it comes to understanding the weak interactions and revealing the internal structure of nucleons and nuclei and has important implications for particle physics and cosmology. Neutrinos are produced abundantly in nature, most notably in circumstances where the weak interactions play a decisive role. These include radioactive decay processes and the nuclear fusion processes that occur inside stars. The measurement of neutrinos produced in the core of the Sun helped establish that the Sun’s energy is a consequence of nuclear fusion dominated by the so-called \texttt{pp} chain.

\subsection{\label{subsec:history_of_nu}Discovering Neutrinos: History and Detection of Neutrino}

For two-body decays, the energy of each of the decay products in the rest frame of the decaying particle is a fixed quantity. Therefore, the observation of continuous electron energy spectra for beta decays of nuclei, where only the daughter nucleus and the emitted electron were observed, posed a puzzle. Either the conservation of energy and momentum was in peril, or something was missing. Pauli proposed (1932) that this missing something is a chargeless and massless particle, termed “neutrino”.

Introducting the hypothetical particle explained both the energy-momentum conservation and the shape of electron spectrum that was observed: 
\begin{eqnarray}
    \dfrac{d \Gamma}{d E_e} \propto p_e E_e\left(E_0 - E_e\right) \sqrt{\left(E_0 - E_e\right)^2 - m_\nu ^2}
\end{eqnarray}
where \(p_e\) and \(E_e\) are momentum and energy of electron and \(E_0 \equiv Q - m_e - m_\nu\) is the maximum energy of electron. 

In the limit when neutrino is massless, the slope of the ``Kurie Plot" is a constant:
\begin{eqnarray}
    \dfrac{d\Gamma / dt}{p_e E_e} \propto \left[(E_0 - E_e)\sqrt{(E_0 - E_e)^2 - m_\nu ^2}\right]
\end{eqnarray}
The beta decay experiments have put an upper limit on electron neutrino mass of \(2.2~\mathrm{eV}\). 

The first direct observation of neutrinos was by Reines and Cowan in 1956\cite{cowan1956detection}, wherein they directed a flux of \(\overline {\nu} _e\) (supposed to have come from a beta decay) into a water target. The reaction \(\overline {\nu} _e p \to n e^+\) produced positrons, which annihilated with electrons in the scintillation counters giving two $0.5 ~\mathrm{MeV}$ photons. The neutrons were absorbed by $\mathrm{CdCl_2}$ dissolved in water, which emitted photons within a few $\mu s$. The coincidence of these two kinds of photons confirmed the above reaction, and hence the presence of \(\overline {\nu} _e\). 

The discovery of \({\nu} _\mu\) took place in 1962 at the Brookhaven National Laboratory \cite{brookhaven} through the decays of pions. Iron from the USS battleship Missouri was used as the target. It was observed that the interactions of these neutrinos with the nuclei (\(\nu_\mu /\overline {\nu}_\mu +N \to \mu _ - /\mu _+ + N' \)) produced only muons, but no electrons. This showed that the ``muon" neutrinos produced in pion decay were distinct from the electron neutrinos produced in beta decays. This was an indication of ``lepton flavour conservation" (which we now know does not hold true in general). The direct observation of \(\nu _\tau\) took place only very recently, in 2000 at the DONUT experiment \cite{KODAMA2001218} at CERN. 
 
\section{\label{sec:neutrino_oscillations}Hints of Neutrino Mass: Neutrino flavours and Neutrino Oscillations}

Like all other fundamental fermions, neutrinos come in \emph{(at least)} three different types, or flavours. A convenient way of describing neutrino production via charged-current weak interactions is to classify the different neutrino flavours according to the flavour of the charged lepton that is produced or destroyed along with the neutrino. Thus, electron-type neutrinos (\(\nu _e\)) are produced and destroyed along with electrons (\(e\)), as are muon-type neutrinos (\(\nu _\mu\)) and tau-type neutrinos (\(\nu _\tau\))  along with muons (\(\mu\)) and taus (\(\tau\)), respectively. 

Since 1998, experiments with neutrinos produced in the Sun, in the atmosphere, in nuclear reactors , and in particle accelerators have demonstrated beyond a reasonable doubt that neutrinos can change flavour as they propagate. The rate of flavour change depends on the neutrino energy, the distance traveled, and the propagation environment. \emph{The only hypothesis capable of explaining all the neutrino data collected during the last few decades is that at least two of the neutrino masses are not zero and are different from one another, and that leptons mix.} In this case, neutrinos change flavour as a function of distance and energy through the phenomenon of neutrino oscillation.\cite{Gouvêa_2016} 

In case of a non-zero rest mass of the neutrino, the flavour  and mass eigenstates are not necessarily identical, a fact well known in the quark sector where both type of states are connected by CKM matrix. This allows for the phenomenon of neutrino oscillations, a kind of flavour oscillations which is already known in other particle systems.  It can be described by
pure quantum field theory. Oscillations are observable as long as the neutrino wave
packets form a coherent superposition of states. Such oscillations among the different
neutrino flavours do not conserve individual flavour lepton numbers, only a total
lepton number.

\subsection{General formalism of Oscillations}

For more detailed version of this discussion [Ref~\onlinecite{zuber2020neutrino}]. Let us assume that there is an arbitrary number of \(n\) orthonormal eigenstates.
The \(n\) flavour eigenstates \(\ket{\nu_\alpha}\) with   \(\bra{\nu_\beta}\ket{\nu_\alpha} = \delta _{\alpha\beta}\) are connected to the \(n\) mass eigenstates \(\ket{\nu_i}\) with \(\bra{\nu_i}\ket{\nu_j} = \delta _{ij}\) via a unitary mixing matrix \(U\):
\begin{gather}
    \ket{\nu _\alpha} = \sum _{i} U_{\alpha i} \ket{\nu _i} \\
    \ket{\nu _i} = \sum _{\alpha} (U^\dagger) _{i \alpha } \ket{\nu _\alpha} = \sum _{\alpha} U^* _{i \alpha } \ket{\nu _\alpha}
\end{gather}

with 
\begin{gather}
    U^\dagger U = 1 \quad \sum_{i} U_{\alpha i} U^* _{\beta i} = \delta _{\alpha\beta}\quad\sum_{\alpha} U_{\alpha i} U^* _{\alpha j} = \delta _{ij}
\end{gather}

For antineutrinos we have:
\begin{equation}
    \ket{\overline {\nu} _\alpha} = \sum _{i} U^* _{\alpha i} \ket{\overline {\nu} _i}
\end{equation}

The mass eigenstates \(\ket{\nu _i}\) are stationary states and show a time dependence according to 
\begin{equation}
    \ket{\nu _i (x, t)} = e^{-i E_i t} \ket{\nu _i (x, o)}
\end{equation}
assuming neutrinos with momentum \(p\) emitted by a source positioned at \(x = 0,(t = 0)\)
\begin{equation}
    \ket{\nu _i (x, 0)} = e^{ip x} \ket{\nu _i}
\end{equation}

and for relativistic case
\begin{eqnarray}
\label{eq:energy_high_energy_approx}
    E_i = \sqrt{m_i ^2 + p_i ^2 } \simeq p_i + \frac{m_i ^2}{2p_i} \simeq E + \frac{m_i ^2}{2E}
\end{eqnarray}
for \(p \gg m_i\) and \(E\approx p\) as neutrino energy. Assume that the difference in mass between two neutrino states with different mass \(\Delta m_{ij} ^2 = m_i ^2 - m_j ^2\) cannot be resolved. Then the flavour neutrino is a coherent superposition of neutrino states with definite mass. Neutrinos are produced and detected as flavour states. Therefore, neutrinos with flavour \(\ket{\nu_\alpha}\) emitted by a source at 
\(t = 0\) propagate with time into a state
\begin{eqnarray}
    \begin{gathered}
\label{eq:neutrino_propogation_p_E}
    \ket{\nu (x, t)} = \sum_{i} U_{\alpha i}e^{-i E_i t}\ket{\nu _i} \\ = \sum_{i, \beta} U_{\alpha i}U^*_{\beta i} e^{ip x}e^{-i E_i t}\ket{\nu _\beta} 
\end{gathered}
\end{eqnarray}
Different neutrino masses imply that the phase factor in (\ref{eq:neutrino_propogation_p_E}) is different. This means
that the flavour content of the final state differs from the initial one. At macroscopic
distances this effect can be large in spite of small differences in neutrino masses. The
time-dependent transition amplitude for a flavour conversion \(\nu_\alpha \to \nu_\beta\) is then given by:
\begin{equation}
\label{eq:transition_amplitude_neutrino}
       A(\alpha \to \beta)(t) = \bra{\nu_\beta}\ket{\nu (x, t)}
       = \sum_{i} U^*_{\beta i}U _{\alpha i} e^{ip x} e^{-i E_i t}
\end{equation}

Using (\ref{eq:energy_high_energy_approx}) this can be written as:
\begin{equation}
    \begin{aligned}
         A(\alpha \to \beta)(t) = \bra{\nu_\beta}\ket{\nu (x, t)} = \sum_{i} U^*_{\beta i}U_{\alpha i} e^{-i \frac{m_i ^2}{2} \frac{L}{E}}\\
         = A(\alpha \to \beta)(L)\\
    \end{aligned}
\end{equation}
with \(L = x = ct\) being the distance between source and detector. In an analogous way, the amplitude for the antineutrino transition can be derived as:
\begin{eqnarray}
\label{eq:transition_amplitude_antineutrino}
    A(\bar{\alpha} \to \bar{\beta})(t) = \sum_{i} U_{\beta i}U^* _{\alpha i} e^{-i E_i t} 
\end{eqnarray}

Hence the transition probability \(P\) can be obtained from the transition amplitude \(A\):
\begin{widetext}
\begin{equation}
\label{eq:transition_probability}
        P(\alpha \to \beta)(t) = \left|A(\alpha \to \beta)(t)\right|^2 
        = \sum_{i}\sum_{j}  U_{\alpha i}U^* _{\alpha j}U^*_{\beta i}U _{\beta j} e^{-i (E_i - E_j) t} 
        = \left|U_{\alpha i}U^*_{\beta i}\right|^2 
        + 2 \mathrm{Re}\sum_{j > i} U_{\alpha i}U^* _{\alpha j}U^*_{\beta i}U _{\beta j} e^{\left(- i \frac{\Delta m_{ij}^2}{2} \frac{L}{E}\right)}
\end{equation}
\end{widetext}
The second term in (\ref{eq:transition_probability}) describes the time (or spatial) dependent neutrino oscillations. The first one is an average transition probability, which also can be written as
\begin{eqnarray}
    \langle{P_{\alpha \to \beta}}\rangle  = \sum_{i}\left|U_{\alpha i}U^*_{\beta i}\right|^2 \nonumber \\
    = \sum_{i} \left|U^*_{\alpha i}U_{\beta i}\right|^2 = \langle{P_{\beta \to \alpha}}\rangle
\end{eqnarray}

Using \textsc{CP} invariance (i.e when \(U_{\alpha i}\) real), the above expression can be simplified to
\begin{widetext}
    \begin{eqnarray}
        P(\alpha \to \beta)(t) 
        = U^2_{\alpha i}U_{\beta i}^2 
        + 2 \sum_{j > i} U_{\alpha i}U _{\alpha j}U_{\beta i}U _{\beta j} \cos{\left( \frac{\Delta m_{ij}^2}{2} \frac{L}{E}\right)}
        = \delta _{\alpha\beta} - 4 \sum_{j > i} U_{\alpha i}U _{\alpha j}U_{\beta i}U _{\beta j} \sin^2 {\left( \frac{\Delta m_{ij}^2}{4} \frac{L}{E}\right)}
\end{eqnarray}
\end{widetext}

The probability of finding the original flavour is given by
\begin{equation}
    P(\alpha \to \alpha) = 1 - \sum _{\alpha \not = \beta} P(\alpha \to \beta)
\end{equation}

As can be seen from (\ref{eq:transition_probability}) there will be oscillatory behaviour as long as at least one neutrino mass eigenstate is different from zero and if there is a mixing (non-diagonal terms in \(U\) ) among the flavours. In addition, the observation of oscillations allows no absolute mass measurement; oscillations are sensitive to only \(\Delta m^2\). Last but not least, neutrino masses should not be exactly degenerated. Another important feature is the dependence of the oscillation probability on $L/E$.

\subsection{CP and \textsc{T} violation in neutrino oscillation}

Notice that (\ref{eq:transition_amplitude_neutrino}) and (\ref{eq:transition_amplitude_antineutrino}) yield a relation between neutrino and antineutrino transition amplitudes:
\begin{equation}
   A(\overline {\alpha} \to \overline {\beta})(t) =  A(\alpha \to \beta)(t)\not = A(\beta \to \alpha)(t)
\end{equation}
This relation is a direct consequence of the \textsc{CPT} theorem. \textsc{CP} violation manifests itself if the oscillation probabilities of \(\nu _{\alpha} \to \nu _\beta\) are different from its \textsc{CP} conjugate process \(\overline {\nu} _{\alpha} \to \overline {\nu} _\beta\). So one observable for detection could be
\begin{equation}
    \Delta P^{CP} _ {\alpha\beta} = P(\nu _\alpha \to \nu _\beta) - P(\overline {\nu} _{\alpha} \to \overline {\nu} _\beta) \not = 0, ~ \alpha \not = \beta
\end{equation}
Similarly, \textsc{T} violation can be tested if the probabilities of  \(\nu _{\alpha} \to \nu _\beta\)
are different from the  \textsc{T} conjugate process  \(\nu _{\beta} \to \nu _\alpha\) . Here, the observable is
\begin{equation}
    \Delta P^{T} _ {\alpha\beta} = P(\nu _\alpha \to \nu _\beta) - P(\nu _{\beta} \to \nu _\alpha) \not = 0, ~ \alpha \not = \beta
\end{equation}
If \textsc{CPT} conservation holds, which is the case for neutrino oscillations in vacuum, violation of \textsc{T}  is equivalent to violation of \textsc{CP}. Using \(U_{PMNS}\) (Eq~\ref{eq:upmns}) it can be shown explicitly that in vacuumm \(\Delta P^{CP} _ {\alpha\beta}\) and \(\Delta P^{T} _ {\alpha\beta}\) are equal and given by \cite{zuber2020neutrino})
\begin{equation}
    \begin{gathered}
        \Delta P^{CP} _ {\alpha\beta} = \Delta P^{T} _ {\alpha\beta}\\
        = - 16 J_{\alpha\beta} \sin \left(\frac{\Delta m^2 _{12}}{4E} L\right)\sin \left(\frac{\Delta m^2 _{23}}{4E} L\right)\sin \left(\frac{\Delta m^2 _{13}}{4E} L\right)
    \end{gathered}
\end{equation}

where \(J_{\alpha\beta} \equiv \mathrm{Im}[U_{\alpha 1}U^*_{\alpha 2}U^*_{\beta 1}U_{\alpha 2}]\). 

\textit{Readers may refer to Ref~\onlinecite{Dighe} (Section 7.3, pg. 31-32) for an alternative approach to same problem}

\subsection{Two state atmospheric neutrino mixing}
In 1998 the Super-Kamiokande experiment measured the number of electron and muon neutrinos that arrive at the Earth’s surface as a result of cosmic ray interactions in the upper atmosphere, which are referred to as ``atmospheric neutrinos". While the number and and angular distribution of electron neutrinos is as expected, it showed that the number of muon neutrinos is significantly smaller than expected gave compelling evidence that muon neutrinos undergo flavour oscillations and this in turn implies that at least one neutrino flavour has a non-zero mass. The standard interpretation is that muon neutrinos are oscillating into tau neutrinos (\textit{the ``electron" neutrinos interact with the atmospheric electron, hence their conversion probability in atmosphere is very low!?}).\cite{King_2004}

The current neutrino atmospheric oscillation data are well described by simple two-state mixing
\begin{equation}
    \begin{pmatrix}\nu _\mu \\ \nu _\tau\end{pmatrix} = \begin{pmatrix}
        \cos \theta_{23} & \sin \theta_{23}\\
        -\sin \theta_{23} & \cos \theta_{23}
    \end{pmatrix}  \begin{pmatrix}
        \nu _2 \\ \nu _3
    \end{pmatrix}
\end{equation}
Using the Effective Hamiltonian from Eq~\ref{eq:energy_high_energy_approx}. Since interference can take place only between neutrinos with the same \(E \approx p\), we shall consider all neutrino fluxes as mixtures of coherent beams. For each coherent beam, the time evolution \(\exp (-iHt)\) contains a common phase \(\exp (-ipt)\), which is irrelevant for oscillations. We shall therefore take the effective neutrino Hamiltonian to be\cite{Dighe}
\begin{equation}
    H_i = \frac{m_i^2}{2E}
\end{equation}
and if neutrinos are produced as a flavour eigenstate \(\nu_\mu\), their time evolution will be:
\begin{equation}
    \begin{gathered}
    \ket{\nu _\mu} = \cos \theta_{23}\ket{\nu _2} + \sin \theta_{23}\ket{\nu _3}\\
     = \cos \theta_{23}e^{- \frac{im_2 ^2 t}{2E}}\ket{\nu _2(0)} + \sin \theta_{23}e^{- \frac{im_3 ^2 t}{2E}}\ket{\nu _3(0)}
    \end{gathered}
\end{equation}
the probability of observing the same flavour eigenstate \(\nu_\mu\) at time \(t\) is 
\begin{equation}
    P_{\mu\mu} = \abs{\braket{\nu_\mu}{\nu_\mu (t)}}^2 = 1 - \sin^2 2\theta_{23}\sin ^2 \left(\frac{\Delta m_{32}^2 L}{4E}\right)
\end{equation}
where \(\Delta m_{32}^2 = m_3 ^2 - m_{2} ^2\). Observe \(P_{\mu\mu} = 1\) at \(t = 0\) and it oscillates  with a depth of \(\sin ^2 2\theta_{23}\) and an oscillation wavelength of \((4\pi E/\Delta m_{32}^2)\).
Hence ``conversion probability" into the other flavour eigenstate \(\nu _\tau\) is 
\begin{equation}
    P_{\mu\tau} = \sin ^2 2 \theta_{23} \sin ^2 \left(\frac{\Delta m_{32}^2 L}{4E}\right)
\end{equation}
The above equation is in `natural" units, where  \(\hbar = c = 1\). If \(\Delta m^2 _{32}\) is in \(\mathrm{eV^2}\), \(L\) in km and \(E\) in GeV, it may be written as
\begin{equation}
    P_{\mu\tau} = \sin ^2 2 \theta_{23} \sin ^2 \left(1.27\frac{\Delta m_{32}^2 L}{4E}\right)
\end{equation}

The atmospheric data is statistically dominated by the Super-Kamiokande results and the latest reported data sample leads to (Ref~\onlinecite{King_2004}):
\begin{equation}
    \begin{gathered}
        \sin^2 2 \theta _{23} > 0.92 \\
    1.3E-3 ~\mathrm{eV}^2 < \abs{m_{32}^2} < 3E-3 ~\mathrm{eV}^2
    \end{gathered}
\end{equation}

\subsection{Three family solar neutrino mixing}
Super-Kamiokande is also sensitive to the electron neutrinos arriving from the Sun, the ``solar neutrinos", and has independently confirmed the reported deficit of such solar neutrinos long reported by other experiments.
Since \(\nu _e\) do not participate in the atmospheric neutrino oscillations, whereas they do participate in the solar neutrino oscillations, it is clear that in order to have a consistent picture of neutrino mixings, we have to develop a framework for the mixing of all three neutrino species, \(\nu _e, \nu _\mu, \nu _\tau \). 

We neglect CP violation for the moment. In the absence of CP violation, the \(3\times 3\) mixing matrix is real and is completely described in terms of the three mixing angles, \(\theta _{12}, \theta _{13},\theta _{23}\). It can be explicitly written as (\textit{called Pontecorvo–Maki–Nakagawa–Sakata matrix}, also \(\delta_{CP} = 0\))

\begin{equation}
\label{eq:upmns}
    \resizebox{0.45\textwidth}{!}{$\begin{gathered}
    U_{PMNS} =  R_{23}R_{13}R_{12} \\
    = {\begin{pmatrix}1&0&0\\0&c_{23}&s_{23}\\0&-s_{23}&c_{23}\end{pmatrix}}{\begin{pmatrix}c_{13}&0&s_{13}e^{-i\delta _{\mathrm {CP} }}\\0&1&0\\ s_{13}e^{i\delta _{\mathrm {CP} }}&0&c_{13}\end{pmatrix}}{\begin{pmatrix}c_{12}&s_{12}&0\\-s_{12}&c_{12}&0\\0&0&1\end{pmatrix}}
    \end{gathered}$}
\end{equation}
The probability that a neutrino flavour eigenstate \(\ket{\nu_\alpha}\) will be converted to another flavour eigenstate \(\ket{\nu _\beta}\) can be calculated as we had seen in the case of two neutrino mixing. In the absence of any matter effect, the probability is given by
\begin{equation}
    \resizebox{0.45\textwidth}{!}{$\begin{gathered}
        P_{\alpha \to \beta} = \delta_{\alpha\beta} - 4 \sum_{i > j} \mathrm{Re}\left(U_{\alpha i}U^* _{\alpha j}U^*_{\beta i}U _{\beta j}\right) \sin ^2 \left(\frac{\Delta m^2 _{ij} L}{4E} \right)\\
        + 4 \sum_{i > j} \mathrm{Im}\left(U_{\alpha i}U^* _{\alpha j}U^*_{\beta i}U _{\beta j}\right) \sin \left(\frac{\Delta m^2 _{ij} L}{4E} \right) \cos \left(\frac{\Delta m^2 _{ij} L}{4E} \right)
    \end{gathered}$}
\end{equation}
With no CP violation, the last term vanishes\cite{Dighe}. The general formulae in the three-flavour scenario are quite complex; therefore, the following assumption is made: in most cases only one mass scale is relevant, i.e., \(\Delta m^2 _{atm} \sim 10^{-3}~\mathrm{eV}^2\). Furthermore, one possible neutrino mass spectrum such as the hierarchical one is taken\cite{zuber2020neutrino}
\begin{equation}
    \Delta m^2 _{21} = \Delta m^2 _{sol} \ll \Delta m^2 _{31} \approx \Delta m^2 _{32} = \Delta m^2 _{atm}
\end{equation}
Then the expressions for specific oscillation transitions are:
\begin{equation}
    \begin{gathered}
        P(\nu _\mu \to \nu _\tau) = 4\left|U_{33}\right|^2\left|U_{23}\right|^2\sin ^2\left(\frac{\Delta m^2 _{atm} L}{4E} \right)\\
        = \sin ^2 (2 \theta _{23})\cos ^2 (\theta _{13})\sin ^2\left(\frac{\Delta m^2 _{atm} L}{4E} \right)
    \end{gathered}
\end{equation}
\begin{equation}
    \begin{gathered}
        P(\nu _e \to \nu _\mu) = 4\left|U_{13}\right|^2\left|U_{23}\right|^2\sin ^2\left(\frac{\Delta m^2 _{atm} L}{4E} \right)\\
        = \sin ^2 (2 \theta _{13})\sin ^2 (\theta _{23})\sin ^2\left(\frac{\Delta m^2 _{atm} L}{4E} \right)
    \end{gathered}
\end{equation}
\begin{equation}
    \begin{gathered}
        P(\nu _e \to \nu _\tau) = 4\left|U_{33}\right|^2\left|U_{13}\right|^2\sin ^2\left(\frac{\Delta m^2 _{atm} L}{4E} \right)\\
        = \sin ^2 (2 \theta _{13})\cos^2 (\theta _{23})\sin ^2\left(\frac{\Delta m^2 _{atm} L}{4E} \right)
    \end{gathered}
\end{equation}

As a side comment, the derivation of the oscillation probability depends on two assumptions: \textit{that
the neutrino flavour and mass states are mixed and that we create a coherent superposition of mass
states at the weak vertex. This coherent superposition reflects the fact that we can’t experimentally
resolve which mass state was created at the vertex. One might ask oneself what we would expect
to see if we did know which mass state was created at the vertex. If we knew that, we would know
the mass of the neutrino state that propagates to the detector. There would be no superposition, no
phase difference and no flavour oscillation. However there would be flavour change.}

\section{\label{sec:mass_in_sm}Neutrinos, Standard Model and Mass Terms}

\begin{figure}[!ht]
    \centering
    \includegraphics[height=0.7\textwidth]{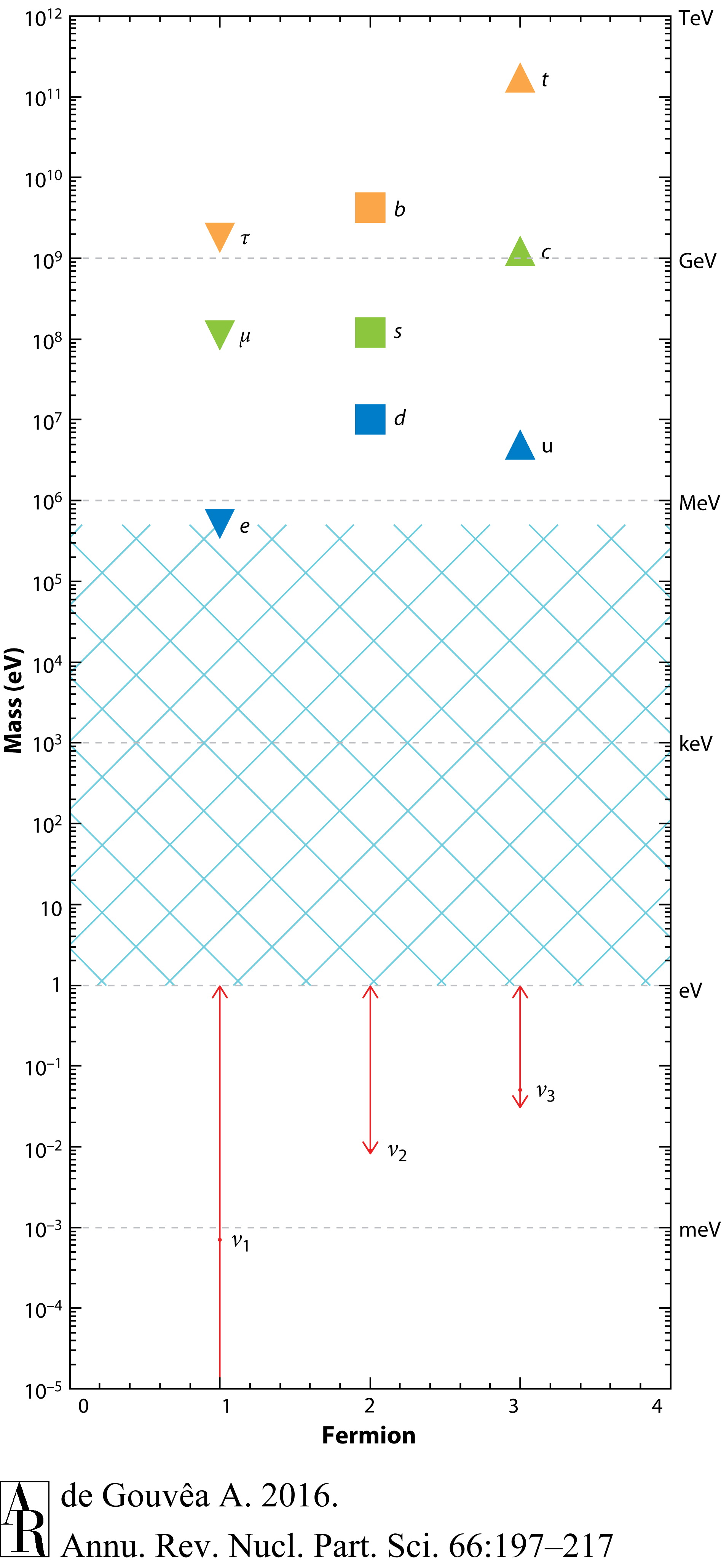}
    \caption{The Standard Model fermion masses.The arrows indicate the allowed ranges for the neutrino masses, assuming a so-called normal mass ordering: $m_{23} ^2> m_{22}^2 > m_{21}^2$ (Ref~\onlinecite{Gouvêa_2016})}
    \label{fig:sm_fremion_mass}
\end{figure}

Although the SM has been marvelously successful for the last couple of decades, in 1998 the breakthrough regarding the neutrino flavour oscillations threw a spanner in the works. The subject of neutrino mass underlies many of the current leading questions in particle physics. Why is there more matter than anti-matter? There are some indications that understanding the mechanism of neutrino mass could help solve this. Why are the neutrinos so light? Figure~\ref{fig:sm_fremion_mass} is a plot of the neutrino mass compared to the mass of the other charged leptons. The neutrino mass is roughly a factor of a million less than the other particles, which only differ in mass by at most a thousand.

We will discuss briefly how some of these questions may be solved.

\subsection{Dirac Mass}
The standard way to include mass in Standard Model is through Dirac Mass Terms in the Lagrangian:
\begin{equation}
        m\overline {\psi} \psi
\end{equation}
If we decompose the dirac spinor \(\psi\) into it's left and right handed-chiral states, we may rewrite the mass term above as
\begin{equation}
    \begin{aligned}
        m\overline {\psi} \psi = m\overline{(\psi_L + \psi _R)}(\psi_L + \psi _R)\\
         = m\overline {\psi} _L \psi_R + m\overline {\psi}_R \psi_L
    \end{aligned}
\end{equation}
above we used the property \(\overline {\psi}_L \psi_L =\overline {\psi}_R \psi_R = 0\).

The important point to note is that a non-zero Dirac mass requires a particle to have both a left-and right-handed chiral state : in fact the \textsc{Dirac mass can be viewed as being the coupling constant
between the two chiral components}. Also, observe this formalism is not gauge invariant.

To begin with, remember that the left-handed neutrino field \(\nu _L\) in the Standard Model is a part of a doublet of a left-handed weak field, other being the left-handed electron: \(\begin{pmatrix}e \\ \nu\end{pmatrix}_L\). These fields are distinguised from each other only by their \emph{weak charge}, also called \emph{weak isospin} \(T_3\). The right-handed particle fields are singlet fields, with no weak charge at all (and hence right-handed particle fields can’t couple to the $W$ bosons, in much the same way as neutrinos can’t couple to photons as they are electrically
neutral

The field \(e_L\) has a weak charge of $T_3 = - \frac{1}{2}$ and an electric charge $Q = -1$. Now, a consequence of being gauge invariant is that all terms in the Lagrangian are neutrally charged; both electrically and from the point of view of the weak interaction. Now consider the Dirac mass term for an electron: $m_e \overline{e}_R e_L$ . The charges on each of the charged lepton fields and the overall term are shown in Table~\ref{tab:charges_sm}.
Note that ``\textit{adjointing}"  a field flips the sign of the charge.
\begin{table}[!h]
    \centering
    \begin{tabular}{|c|c|c|c|c|}\hline
    & $e_L$ & $e_R$ & $\overline {e}_R$ & $\overline {e}_R {e}_L$\\\hline
    $Q$ & $-1$ & $-1$ & $+1$ & $0$\\
    $T_3$ & $-1/2$ & $0$ & $0$ & $-1/2$  \\\hline
    \end{tabular}
    \caption{Charges of fields in the electron Dirac mass term}
    \label{tab:charges_sm}
\end{table}
Since the dirac mass terms are not gauge invariant, we need to add a field to mass term which (i) has \(Q = 0\) (ii) has \(T_3 = 1/2\). This field would be electrically neutral but part of weak doublet, just like the left-handed electron field. This field will turn out to be the neutral Higgs boson. With this addition, the Dirac mass term becomes neutrally charged and the term is gauge invariant.

Since we know the neutrino is chirally left-handed, there is good reason to suppose that the neutrino must be massless, as there does not seem to be a right-handed state for the mass term to couple to. However, we now know that the neutrino does have a small mass, so either there must be a right-handed neutrino which only shows up in the standard model to give the neutrino mass, but otherwise cannot be observed as the weak interaction doesn’t couple to it, or there is some other sort  of mass term out there.

By definition Dirac particles come in four distinct types: left- and right-handed particles, and left- and right-handed antiparticles, and are therefore represented by a 4-component Dirac spinor. As we will see in the next section, charged fermions can only have a Dirac type mass. For neutral fermions, however, this need not be the case. Such particles could acquire mass through the \emph{Majorana mass mechanism}.

\subsection{\label{sub:sm_mass_models}The Majorana mechanism}
We have seen that the Dirac mass mechanism requires the existance of a sterile right-handed neutrino state. In the early 1930’s, a young physicist by the name of Ettore Majorana wondered if he could dispense with this requirement and construct a mass term using only the left-handed chiral state
Let us first split the Dirac Lagrangian up into its chiral components
\begin{eqnarray}
    \begin{gathered}
         \mathcal{L} = \overline {\psi} (i \gamma _\mu \partial^\mu - m)\psi\\
         = (\overline {\psi}_L + \overline {\psi}_R) (i \gamma _\mu \partial^\mu - m)(\psi_L + \psi_R)\\
         = \overline {\psi}_L (i \gamma _\mu \partial^\mu - m\overline {\psi}_R)\psi_L+\overline {\psi}_R (i \gamma _\mu \partial^\mu - m\overline {\psi}_L)\psi_R
    \end{gathered}
\end{eqnarray}
we have used the fact (Appendix~\ref{sec:charge_conjugation}) that \(\overline {\psi}_R \gamma ^\mu\partial _\mu {\psi}_L = \overline {\psi}_L \gamma ^\mu\partial _\mu {\psi}_R = 0\) and \(\overline {\psi}_R {\psi}_R = \overline {\psi}_L {\psi}_L = 0\)

Using the Euler Lagrange equation we can look for independent equations of motion for the left-and right-handed fields, \(\psi _R\) and \(\psi_L\) . We obtain two coupled Dirac Equations
\begin{equation}
\label{eq:coupled_dirac_eq}
    \begin{gathered}
        i \gamma ^\mu \partial_\mu \psi _L = m \psi _R\\
        i \gamma ^\mu \partial_\mu \psi _R = m \psi _L
    \end{gathered}
\end{equation}
The mass term couples the two equations. For massless fields; we have the Weyl equations
\begin{equation}
    \begin{gathered}
        i \gamma ^\mu \partial_\mu \psi _L = 0\\
        i \gamma ^\mu \partial_\mu \psi _R = 0
    \end{gathered}
\end{equation}
The neutrino is now described using only two independent two-component spinors which turn out to be helicity eigenstates and describe two states with definite and opposite helicity. These correspond to the left-handed neutrino and the right-handed neutrino. Since the right-handed neutrino field does not exist, then we just describe the neutrino with a single massless left-handed field and that is that. This is the usual formulation of the Standard Model.

Ettore Majorana wondered if he could describe a massive neutrino using just a single left-handed field. At first glance this is impossible as you need the right-handed field to construct a Dirac mass term. Majorana, however, found a way.
Let’s take the second equation in Eq~\ref{eq:coupled_dirac_eq}. We want to try to make it look like the first by finding an expression for \(\psi _R\) in terms of \(\psi_L\) . To begin with, let’s take the hermitian conjugate of the second equation
\begin{equation}
\label{eq:weyl_eq_massive_neutrino}
    \begin{gathered}
        (i \gamma ^\mu \partial_\mu \psi _R)^\dagger = (m \psi _L)^\dagger\\
        -i \partial_\mu \psi _R ^\dagger (\gamma ^\mu) ^\dagger = m \psi _L ^\dagger
    \end{gathered}
\end{equation}
Multiplying on the right by \(\gamma^0\) we get
\begin{equation}
    -i \partial_\mu \psi _R ^\dagger(\gamma ^\mu) ^\dagger \gamma^0 = m \psi _L ^\dagger\gamma^0
\end{equation}
One of the properties of the \(\gamma\) matrices is \(\gamma^0 (\gamma^\mu) ^\dagger\gamma^0 = - \gamma_\mu\) . Multiplying on the left by \(\gamma^0\) and remembering that \((\gamma^0)^2 = 1\), we have \((\gamma^\mu) ^\dagger\gamma^0 = \gamma^0\gamma^\mu\). Using this
\begin{equation}
    -i \partial_\mu \psi _R ^\dagger \gamma^0 \gamma^\mu = m \psi _L ^\dagger\gamma^0
\end{equation}
and therefore 
\begin{equation}
    -i \partial_\mu \overline {\psi} _R \gamma ^\mu = m \overline {\psi}_L
\end{equation}

We want this to have the same structure as the first equation (Eq~\ref{eq:weyl_eq_massive_neutrino}), but that negative sign out the front and the wrong position of the \(\gamma^\mu\) matrix is spoiling this. We can deal with this by taking the transpose
\begin{equation}
    \begin{gathered}
        - i [\partial _ \mu \overline {\psi}_R \gamma^\mu]^T = m \overline {\psi}_L ^T\\
        - i (\gamma^\mu)^T \partial _ \mu \overline {\psi}_R ^T = m \overline {\psi}_L ^T
    \end{gathered}
\end{equation}
and using the property of the charge conjugation matrix that \(C (\gamma^\mu)^T = - \gamma^\mu C\) we get
\begin{equation}
    i\gamma^\mu \partial_\mu C \overline {\psi}_R ^T = m \overline C{\psi}_L ^T
\end{equation}
This equation has the same structure as the first if we require that the right handed component of \(\psi\) is
\begin{equation}
    \psi _R = C \overline {\psi}_L ^T
\end{equation}
This assumption requires that \(C \overline {\psi}_L ^T\) is actually right-handed. Is this true? Well, if the field is right-handed then applying the left-handed chiral projection operator, \(P_L = \frac{1}{2} (1 - \gamma^5); P_L\psi _R = 0\). Using the properties of the charge conjugation matrix, \(P_L C = C P_L ^T\), we have
\begin{equation}
    P_L (C \overline {\psi}_L ^T) = C P_L ^T  \overline {\psi}_L ^T = C (\overline {\psi}_L P_L)^T
\end{equation}
Now
\begin{equation}
    \begin{aligned}
        \overline {\psi}_L P_L = (P_L\psi)^\dagger \gamma_0 P_L\\
        = \psi ^\dagger P_L \gamma_0 P_L
        = \psi ^\dagger \gamma_0 P_R P_L= 0\\
    \end{aligned}
\end{equation}
So, yes, if we define the right-handed field \(\psi _R = C \overline {\psi}_L ^T\), then we can write the Dirac equation only in terms of the left-handed field \(\psi _L\). The Majorana field, then becomes
\begin{equation}
\psi = \psi _L + \psi _R = \psi _L + C \overline {\psi}_L^T = \psi _L + \psi _L ^C
\end{equation}
where we've defined the \textit{charge-conjugate field}, \(\psi _L ^C = C \overline {\psi}_L^T\)
What does this imply? Well, let's take the charge conjugate of the Majorana field:
\begin{equation}
    \psi ^C = (\psi _L + \psi _L ^C)^C = \psi _L ^C + \psi _L = \psi
\end{equation}
That is, the charge conjugate of the field is the same as the field itself, or more prosaically, a Majorana particle is it's own anti-particle.

What sort of particle can be Majorana - well, clearly it must be neutral, as the charge conjugation operator flips the sign of the electric charge. Any charged fermion therefore will not be identical to its antiparticle. In fact, the only neutral fermion that could be a Majorana particle is the neutrino.

This also changes our view of the nature of the neutrino. To date we have been assuming that the neutrino and the antineutrino are distinct particles, but let’s think about that. We define the neutrino to be that left-handed state which is created together with a negatively charged lepton in the decay of a \(W^-\) boson. and the antineutrino to be that right-handed state which is created together with a positively charged lepton in the decay of \(W^+\) boson. However, we never see the neutrino or antineutrino particles themselves. If the neutrino was a Majorana particle, with a left- and right - handed component then we would only have a single particle: the neutrino. When a \(W^-\) boson decayed it would naturally produce the left-handed component as that’s the only bit of the neutrino field which couples to the \(W^-\) , and vice-verse if a \(W^+\) decays. We don't need two independent particles- just one with two independent chiral components will do.

\subsubsection{Majorana Mass Term}
We saw above that the mass term in the Lagrangian couples left- and right-handed neutrino chiral states: \(\mathcal{L}^D = - m \overline {\nu}_R \nu _L\). If the particle is Majorana, We can also form a mass term just with the left-handed component. In this case, the right-handed component is \(\nu _L ^C = C \overline {\nu}_L^T\)
\begin{equation}
    \mathcal{L}^M _L = -\frac{1}{2} m \overline {\nu ^C _L} \nu _L
\end{equation}
The factor of a half there is to account for double-counting since the hermitian conjugate is identical.
\subsubsection{Lepton Number violation} 
The Majorana term couples the antineutrino to the neutrino component. Dirac neutrinos have lepton number $L = +1$ and antineutrinos have lepton number $L = -1$. Since Majorana neutrinos are the same as their antiparticle it is impossible to give such an object a conserved lepton number. Indeed, interactions involving Majorana neutrinos generally violate lepton number conservation by \(\Delta L = \pm 2\).

\subsubsection{$^\ast$Neutrinoless double beta decay}

The experiments involving neutrino oscillations cannot identify whether neutrino is a Majorana particle. However, if we observe neutrinoless double beta decay, would confirm what the Majorana nature of neutrino is the:
\begin{equation}
    N (A, Z) \to N '(A, Z + 2) + 2e^-
\end{equation}
where the nucleus $N$ with atomic mass number $A$ and atomic number $Z$ decays to the nucleus $N'$ with atomic mass number $A$ and atomic mass number $Z + 2$ without any neutrino emission (i.e. without any missing energy). Such a reaction is characterized by the two electrons emitted back to back with the same energy, the combined energy being exactly equal to the $Q$ value of the reaction.

The cross section for this process is proportional to a particular combination of Majorana masses:
\begin{equation}
    \sigma \propto \abs{\sum U^2 _{ei} m_i}^2
\end{equation}
so that the number of events observed are a measure of the ``effective" Majorana mass of the electron neutrino:
\begin{equation}
    \langle m_{ee} \rangle = \abs{\sum U^2 _{ei} m_i}
\end{equation}
Till now no confirmed neutrinoless double beta decay events have been observed, which has enabled us to put an upper bound of \(\langle m_{ee} \rangle < 0.35~\mathrm{eV}\).\cite{Aalseth:2004hb}

\section{\label{sec:mass_mechanism}Neutrino mass mechanisms}
An enormous amount of literature exists regarding models of neutrino mass generation from grand unified theories, or by adding new particles that interact with neutrinos giving them masses. In the context of this term paper, we will describe some of these different mechanisms that could generate neutrino masses. These include the seesaw mechanism and radiative corrections. There are several different kinds of see-saw mechanism in the literature. In this review we shall focus on the simplest Type I see-saw mechanism, which we shall introduce below. We shall also briefly discuss the type II see-saw mechanism and mention the radiative corrections. For more detail on other neutrino mass models reader may refer to [\onlinecite{Dighe, King_2004, ma_arxiv}]
\subsection{\label{subsec:seesaw_mechanism}Seesaw Mechanism}

We've seen that if only the left-handed chiral field, \(\nu _L\) , exists then there can be no Dirac-type mass term. In this case the neutrino Lagrangian can contain the Majorana mass term
\begin{equation}
    \mathcal{L}^L _{Maj} = - \frac{1}{2}m_L \overline{\nu _L ^C} \nu _L + h.c.
\end{equation}
and the neutrino is a Majorana particle. Unfortunately, given the build-up we've been setting up, such a term cannot actually exist in the standard model, and so \(m_L = 0\). The reason comes from the Higgs mechanism. Recalling the discussion of the Dirac mass term for the electron above, we can write down a table of charges for the left-handed Majorana mass term:
\begin{table}[!h]
    \centering
    \begin{tabular}{|c|c|c|c|c|}\hline
    & $\nu_L$ & $\nu_L ^C$ & $\overline{\nu_L ^C}$ & $\overline{\nu_L ^C}{\nu}_L$\\\hline
    $Q$ & $0$ & $0$ & $0$ & $0$\\
    $T_3$ & $+1/2$ & $-1/2$ & $+1/2$ & $1$  \\\hline
    \end{tabular}
    \caption{Charges of fields in the left-handed Majorana mass term}
    \label{tab:majorana_charges_sm}
\end{table}
Clearly, whilst we've solved the probably of introducing a right-handed neutrino field into the Standard Model, it is still composed of a left-handed field and therefore still carries the quantum numbers of the left-handed field, including non-zero weak isospin. In order to make the left-handed Majorana mass term gauge invariant you therefore need to introduce a field with $Q = 0$ and $T_3 = -1$. This field cannot be part of a Higgs doublet - rather it is part of a Higgs triplet - a set of three fields distinguished by values of $T_3 = -1, 0 ~\text{and}~ 1$. The Standard Model does not include Higgs triplets -and the data does not seem to show any evidence of one. Amongst other things, such a model would allow lepton flavour violating interactions like \(\mu \to e\gamma\) which have never been observed. A left-handed mass term like \(\overline{\nu _L ^C} \nu _L\) doesn’t seem to be possible in the current Standard Model.

Hence, whatever one does, that the neutrino has a mass must imply either (i) there is something really odd we haven’t thought of (never discount this possibility) or (ii) a right-handed chiral neutrino field exists that only interacts with gravity and the Higgs mechanism.

We’ll set $m_L$ equal to zero below, but let’s first assume that we can write a left-handed Majorana mass term \(\mathcal{L}^M _L = 1/2 m_L \overline{\nu_L^c}\nu_L + h.c\). If we’re resigned to have a right-handed neutrino field as well, which we’ll call $N_R$ , we can write a Dirac mass term $\mathcal{L}_D = N_R \nu _L + h.c$. and a further right-handed Majorana field \(\mathcal{L}_R ^M = 1/2 m_R \overline{N_R ^C}N_R +h.c\). We also have the charge-conjugate fields, $\nu_L$ and $N_R$ . These can form another Dirac mass term, $m_D \overline{\nu _L ^C} N_R^C$. The mass for this term must be the same as for the other Dirac mass term, as the total Majorana fields are \(\nu _L ^C\) and \(N_R^C\) , and \(N_R^C +N_R\).

\paragraph*{Combining Majorana and Dirac Mass:} In general, both the Dirac and Majorana mass terms may be present in the Lagrangian of a theory. Of course, this \textit{implies that the theory has right handed neutrinos}, and also incorporates \textit{lepton number violation}. We shall see one example, which can give us some insight into the mechanism of neutrino mass generation.. If one includes all these terms, then the most general mass term one can write down is
\begin{equation}
    \begin{aligned}
        2 \mathcal{L}_{mass} = m_D \overline{N_R}\nu _L + m_D \overline{\nu _L ^C} N_R^C + m_L \overline{\nu_L^C}\nu_L \\
        + m_R \overline{N_R ^C}N_R + h.c.
    \end{aligned}
\end{equation}
which when written as matric equation 
\begin{equation}
    \mathcal{L}_{mass} \sim \begin{pmatrix}\overline{\nu_L^C} & \overline{N_R}\end{pmatrix} \begin{pmatrix}
        m_L & m_D\\
        m_D & m_R
    \end{pmatrix}\begin{pmatrix}
        \nu_L\\N_R ^C
    \end{pmatrix} + h.c.
\end{equation}
the first row vector \(\begin{pmatrix}\overline{\nu_L^C} & \overline{N_R}\end{pmatrix}\) has right-handed fields, the mass matrix is 
\begin{equation}
\label{eq:mass_matrix}
    \mathcal{M}  = \begin{pmatrix}
        m_L & m_D\\
        m_D & m_R
    \end{pmatrix}
\end{equation}
We have expressed the mass Lagrangian in terms of the chiral fields : \(\nu _L\) and \(N_R\) . These fields clearly do not have a definite mass because of the existence of the off-diagonal term, \(m_D\) , in the mass matrix. That means that these fields are not the mass eigenstates, and do not correspond to the physical particle - which must have a definite mass. The values of $m_D$ and $m_L$ are just numbers which indicate how strong the fields in each term couple to each other - they
don’t represent physical masses. This should be no concern - it just means that the flavour eigenstate which couples to the $W$ and $Z$ bosons is a superposition of the massive neutrino states.

In order to find the mass of these states, we need to rewrite the lagrangian in terms of mass eigen states \(\nu _1, \nu_2\). The eigen values of the mass matrix (Eq~\ref{eq:mass_matrix}) above gives the mass of the neutrinos. This matrix has eigenvalues of
\begin{equation}
    \begin{gathered}
        m_1 = \frac{1}{2}\sqrt{4m_D^2 + (m_R - m_L)^2} - \frac{m_R+m_L}{2}\\
        m_2 = \frac{1}{2}\sqrt{4m_D^2 + (m_R - m_L)^2} + \frac{m_R+m_L}{2}
    \end{gathered}
\end{equation}
We can choose different values for $m_L$ ,$m_R$ and $m_D$ and this will give us different physical masses. But a few extreme cases are interesting. 

If one chooses $m_R \gg m_D, m_L$, also called the ``\emph{seesaw}" scenario. \textbf{The special case when \(m_L = 0\) is the ``\textsc{Type}   \texttt{I}" seesaw}, it was was introduced in 1980 (Ref~\onlinecite{MohapatraPhysRevLett}) to provide an explanation for the smallness of neutrino masses as a direct consequence of the heaviness of right-handed neutrinos. As we have seen, the standard model explicitly forbids the left-handed Majorana term ($m_L = 0$) (Table~\ref{tab:majorana_charges_sm}) but says nothing about the right-handed Majorana term, so this choice of parameters is sensible. If we makes this choice, we get the mass of the \(\nu_1\) field to be
\begin{equation}
    m_1 \approx \frac{m_D ^2}{m_R}
\end{equation}
mass of the \(\nu_2\) field becomes 
\begin{equation}
    m_2 = m_R \left(1 + \frac{m_D ^2}{m_R^2}\right)\approx m_R
\end{equation}
Note, notice what happens if we set the mass of the unseen right-handed fields, $m_R$ , to something very big. Then, the physical neutrino with mass
$m_2$ also acquires a very large mass, since $m_2 \approx m_R$.
The physical mass of the other neutrino $m_1$, on the other hand, becomes very small as it is suppressed by the factor of $\frac{1}{m_R}$.

If we try to find expressions for our mass eigenstates, we find that if $m_R$ is very large, \(\nu _1 \sim (\nu _L + \nu -L^C) - \frac{m_D}{m_R ^2}(N_R + N_R^C)\) and \(\nu _2 \sim (N_R + N_R^C)  + \frac{m_D}{m_R ^2}(\nu _L + \nu -L^C)\), i.e.  $\nu _1$ with small mass $m_1$ , is mostly our familiar left-handed light Majorana neutrino and \(\nu _2\) with very heavy mass $m_2$ is mostly the heavy sterile right-handed partner.

This is the famous see-saw mechanism. It provides an explanation for the question of why the neutrino has a mass so much smaller than the other charged leptons. The charged leptons are Dirac particles and therefore have a Dirac mass on the order of $1~\mathrm{MeV}$ (or so). Suppose the Dirac mass of the neutrino is around the same value ($m_D\approx 1~\mathrm{MeV}$) like all the other particles. Then if the mass of the heavy partner is around $10^{15}~\mathrm{eV}$, the mass of the light neutrino will be in the $~\mathrm{meV}$ range, as we now know it is.\cite{boyd}

This is the only natural explanation of the relative smallness of the neutrino mass we currently have. It requires that the neutrino be a Majorana particle, and that there exists an extremely heavy partner to the neutrino with a mass too large for us to be able to create it. This may seem a bit of a stretch, but we know that such particles would have been created very early in the universe. They no longer exist as stable particles, as they have decayed to lighter states as the universe cooled, but due to the uncertainty principle, could exist for the short time necessary to generate mass.

\subsection{Type II Seesaw Mechanism}
What if the neutrino mass originated from another VEV (vacumm expectation value), not the one from the Higgs but one coming from another scalar? The seesaw model type II, introduced in 1981\cite{type3seesaw}, is a tentative to build an answer to that question. Indeed, instead
of extending the fermion content, we can extend the SM with a new heavy scalar which will couple to the neutrino, providing it with a mass
The idea is to add to the SM Lagrangian a new scalar field \(\xi\) ,an $SU(2)_L$ triplet with hypercharge -1($\xi$ is usually referred to as a Higgs boson triplet). This scalar triplet of form: $\Xi \sim \left(\xi ^{++}, \xi ^{+} ,\xi^{0}\right)$\footnote{Charge conservation rules out the possibility for a singlet.} It can also be expressed in terms of the charge eigenstates, then composed of a doubly charged, a singly charged, and a neutral component:
\begin{equation}
    \bm{\Xi} = \begin{pmatrix}
        \xi ^{++} \\ \xi ^{+} \\\xi^{0}
    \end{pmatrix} \equiv \begin{pmatrix}
        \frac{\xi^{1} - i \xi ^{2}}{\sqrt{2}} \\ \xi^{3} \\
        \frac{\xi^{1} + i \xi ^{2}}{\sqrt{2}}
    \end{pmatrix} 
\end{equation}
This scalar triplet can couple to fermions, and the yukawa interactions with leptons can be written as:
\begin{equation}
    \mathcal{L}_{\Xi, Yuk} = - \frac{Y_{\Xi}}{\sqrt{2}} \overline{l^c _L}^f (\bm{\sigma}
    \cdot\bm{\Xi}) l_L ^g + h.c.
\end{equation}
Regarding the scalar potential, the most general potential is\cite{bouchandthesis}:
\begin{equation}
   \begin{gathered}
       V = m^2 \bm{\Phi}^\dagger \bm{\Phi} + M^2 \Xi^\dagger \Xi + \frac{\lambda_1}{2}(\Xi^\dagger \Xi)^2 + \lambda _3 (\bm{\Phi}^\dagger \bm{\Phi})(\Xi^\dagger \Xi) \\+ \mu \left(
        \overline{\xi^{0}} \phi^0 \phi^0 + \sqrt{2}\xi^{-}\phi^{+}\phi^{0} + \xi^{--}\phi^{+}\phi^{+}\right) 
        + h.c.
   \end{gathered}
\end{equation}
Besides, similarly to the Higgs mechanism, in order to get a mass term the scalar triplet is given a VEV breaking the symmetries. The requirement for charge conservation fixes uniquely the breaking direction and therefore it is the neutral component which will acquire a VEV. \textit{The scalar triplet VEV has to be extremely small compared to the Higgs VEV, to account for the tiny mass of neutrinos}. 
\begin{equation}
    \langle \xi^0\rangle = u \ll v = \langle \phi^0 \rangle
\end{equation}
After the symmetry breaking we get a mass term for the neutrinos:
\begin{equation}
    \mathcal{L}_{\Xi, Yuk} =  Y_{\Xi} \overline{\nu^c _L}^f \langle \xi^0\rangle \nu_L  + h.c.
\end{equation}
Thus, using the expression for the VEV that can be calculated from the scalar
potential, it yields
\begin{equation}
    \mathcal{M}_\nu = - uY_{\Xi} \sim +v^2 \mu \frac{Y_{\Xi}}{M^2}
\end{equation}
In this expression one can notice that the light neutrino mass matrix is inversely
proportional to the mass of the scalar triplet, \textsc{consequently this model can also be seen as a `Seesaw'.}

\subsection{Radiative Corrections: Overview of Some Famous Radiative Models}

In the context of the Glashow-Weinberg-Salam theory of electroweak interactions it is considered that all the fermions (except the neutrinos) acquire mass through Yukawa couplings with the Higgs boson after spontaneous symmetry breaking. However, in a radiative model this term is forbidden by some symmetry and there is no mass at tree-level, i.e. the zero order mass vanishes, but the mass is generated at loop-level, at \textit{higher order} in the perturbative expansion. In this section we overview some of the famous Radiative collections to neutino masses. 

\subsubsection{Zee-Wolfenstein Model}
\begin{figure}[!htb]
    \centering
    \includegraphics[scale=0.34]{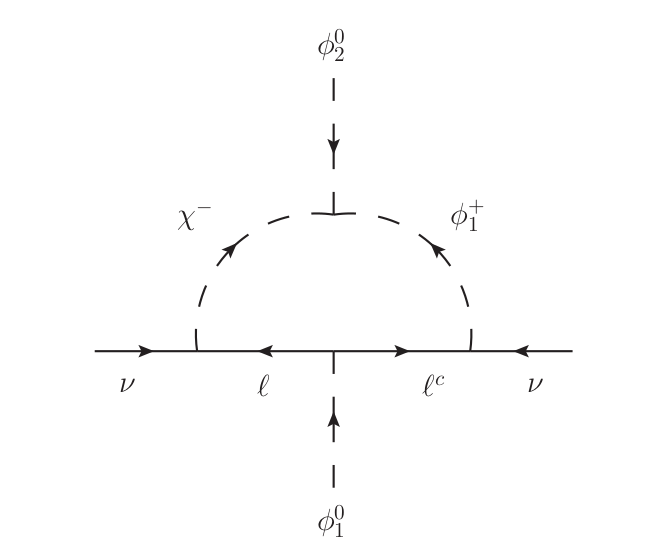}
    \label{fig:zee_wofl}
\caption[]{Zee-Wolfenstein one-loop neutrino mass generation}
\end{figure}

One of the most famous radiative model is the so called Zee-Wolfenstein model\cite{zee_wolfstein}. It extends the scalar sector of the SM with two additional scalars: one charged singlet, \(\chi^+\) and one doublet \(\phi_2 = \begin{pmatrix}
    \phi ^+ _2 \\ \phi _2 ^0
\end{pmatrix}\) (the SM Higgs doublet remains). It uses the fact that the  invariant combination of two (different) lepton doublets couples to a  charged scalar singlet, i.e. $(\nu_i l_j - l_i \nu_j) \chi^+$.  By the same  token, two different scalar doublets are also needed, i.e. $(\phi_1^+ \phi_2^0 - \phi_1^0 \phi_2^+) \chi^-$. Moreover \(\phi_2\) is assumed not to couple to leptons. In this model it is not possible to have tree-level neutrino masses from a renormalizeable Lagrangian, but it is possible at one loop level. Such one-loop process is shown in Fig.~\ref{fig:zee_wofl}. Furthermore, the interesting feature of this model is that it suggests a special form for the neutrino mass matrix:

\begin{equation}
\resizebox{0.46\textwidth}{!}{${\cal M}_\mu = \begin{pmatrix}0 & f_{\mu e} (m_\mu^2 - m_e^2) & f_{\tau e} 
(m_\tau^2 - m_e^2) \\  f_{\mu e} (m_\mu^2 - m_e^2) & 0 & f_{\tau \mu} (m_\tau^2 
- m_\mu^2) \\  f_{\tau e} (m_\tau^2 - m_e^2) & f_{\tau \mu} (m_\tau^2 - m_\mu^2) 
& 0 \end{pmatrix}$}
\end{equation}
This model was studied intensively, but it is now ruled out by data\cite{He:2003ih}.

\subsubsection{Zee-Babu Model}

\begin{figure}[!htb]
    \centering
    \includegraphics[scale=0.3]{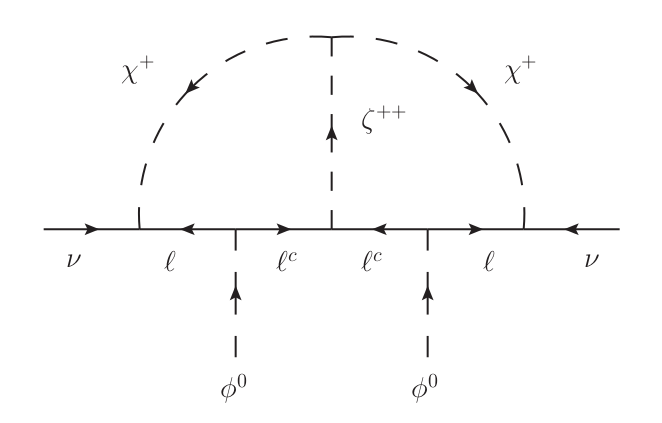}
    \label{fig:zee_babu}
\caption[]{Two-loop neutrino mass generation in the Zee-Babu model}
\end{figure}

In this model introduced between 1985 and 1987~\cite{Zee:1985id, Babu:1988ki} , the Standard Model is
extended to include two charged singlet Higgs fields: a singly charged $\chi ^+$ and a doubly charged $\xi ^{++}$ , while the right-handed neutrinos are not introduced. The outcome is that the resulting neutrino masses are finite and naturally small, since they arise via two-loop diagrams (see Fig.\ref{fig:zee_babu}). The main prediction of the model is the masslessness of one of the neutrinos but it also predicts that some lepton
number changing decays such as $\mu \to eee$ and $\tau \to \mu \mu \mu$ occur at tree level via $\xi ^{++}$ exchange, which provides a tool for its discovery (as well as it acts as a constraint).

\subsubsection{Generic 2-W Mechanism in the SM}
\begin{figure}[!htb]
    \centering
    \includegraphics[scale=0.3]{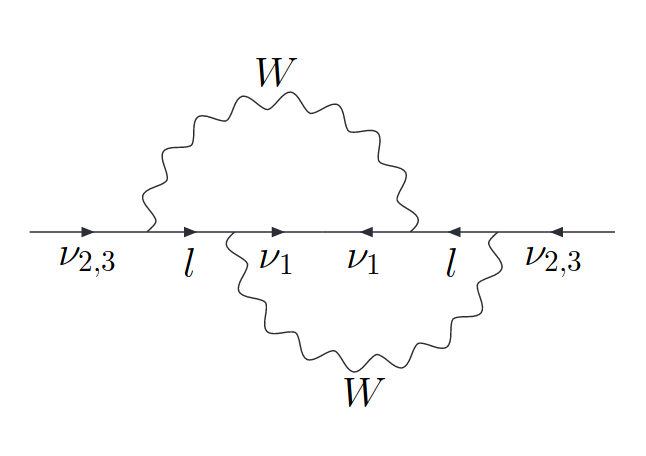}
    \label{fig:two_W}
\caption[]{Two-loop neutrino mass generation in the Zee-Babu model}
\end{figure}

As discussed in Sec.~\ref{sub:sm_mass_models}, the minimal model providing mass for the neutrinos is to add just one neutrino singlet $N$. In that simple case, only one linear combination of the flavour eigenstates ($\nu _{e \mu \tau}$) gets a tree-level Majorana mass term while the two others seem to remain massless. However, these zeros are not protected by any symmetry and nothing forces that it remain as such after radiative corrections. And actually as we can see in Fig.~\ref{fig:two_W} there is a two-loop correction generating mass for the other neutrinos. 

This model introduced in 1988 \cite{babu_ma_phyrev} was formally interesting because although not being totally a radiative model (as one neutrino does have a tree-level mass term) it showed the possibility for the $W$-boson (discovered only five years earlier) to intervene in the generation of neutrino mass only at higher order. However, this two-loop and doubly Glashow-Ilioupoulos-Maiani suppressed mechanism yields completely unsignificant masses for the relevant neutrinos.

\section{Implication of Neutrino Mass,  Limitations and Open Questions}

Neutrino masses are known to be nonzero. They are also tiny when compared with all other mass scales in the SM. Non-zero neutrino masses are bonafide evidence for \textit{physics beyond the SM}. New ingredients must be added to the our current understanding of particle physics in order to explain the phenomena revealed by neutrino oscillation experiments. But there are a lot of questions yet to be answered and understood and knowing the correct answer to these questions could lead to paradigm-shifting developments in physics and astrophysics.  Example: The unitarity of the transformation connecting the mass eigenstates to the electroweak eigenstates is not firmly established. There are tantalizing hints, but no firm evidence, of the existence of sterile neutrinos that do not couple to the vector bosons of the Standard Model, but nevertheless mix with the neutrinos that do. We do not know the transformation properties of the neutrinos under particle-antiparticle conjugation (i.e. whether the neutrinos are Majorana or Dirac fermions). We have just started exploring the full potential of the neutrinos in astrophysics and cosmology. We have just glimpsed into the whole new world of neutrinos, a new domain of physics which stands waiting to be explored!

Following what we discusses in earlier section, the following possibilities arise. Either: 
\begin{itemize}
    \item the neutrino is a Dirac particle, and hence there must be a right-handed chiral neutrino state, which does not interact with matter. This is a so-called \textit{sterile neutrino}.
    \item the neutrino is a Majorana particle. In this case, the neutrino and the antineutrino are identical. The mass term directly couples the left-handed neutrino with the right-handed antineutrino, which implies that the neutrino must have mass. Further, such an interaction implies that lepton number is violated by 2.
    \item If the neutrino is Majorana, then it is possible to explain the very low value of the neutrino mass compared to the quark and charged lepton masses at the expense of the introduction of another very heavy Majorana neutrino though the see-saw mechanism. \textsc{CP} violation in the decays of this heavy neutrino in the early universe could have created the baryon asymmetry we see today.
\end{itemize}

The possibility of the existence of heavy neutrinos, as discussed in  Section ~\ref{subsec:seesaw_mechanism}, also have given rise to another intriguing idea called \textit{leptogenesis} which is intrinsically linked to the question of the matter-antimatter asymmetry. The idea is that these very heavy neutrinos, which are Majorana particles, decayed as the universe cooled into lighter left-handed neutrinos or right-handed antineutrinos, along with Higgs bosons, which themselves decayed to quarks. If the probability of one of these heavy neutrinos to decay to a left-handed neutrino was slightly different than the probability to decay to a right-handed anti-neutrino, then there would be a greater probability to create quarks than anti-quarks and the universe would be matter dominated. More formally, it is thought that the quantum number B - L, where B is the baryon number of the universe and L is the lepton number, must be conserved. If there was some violation of L in the decays of the heavy Majorana neutrinos, this would manifest as a violation in B, and hence the missing anti-matter problem could actually arise from CP violation in the neutrinos. Although there is no direct connection between CP violation in the heavy neutrinos and CP violation in the light neutrinos, this idea still motivates the current attempt to measure CP violation in the light neutrinos at today’s long baseline neutrino experiments.

\section{Conclusion}
We elucidated various mechanisms to generate non-zero neutrino mass. We begin with a discussion on the history of neutrinos and their discovery in Section~\ref{sec:intro}. In Section~\ref{sec:neutrino_oscillations}, we introduce neutrino oscillation and develop the general mathematical formalism for neutrino oscillation, applying it to two flavor and three flavor mixing in the cases of atmospheric and solar neutrinos, respectively. Section~\ref{sec:mass_in_sm} provides a comprehensive description of various mass terms in the Standard Model and their implications. Finally, in Section~\ref{sec:mass_mechanism}, we discuss various mass mechanisms to generate neutrino mass, including Type I and Type II seesaw mechanisms, as well as radiative corrections. There is a vast body of literature on Neutrino Physics and open questions in neutrino physics, and this term paper covers only some aspects of it. Interested readers are encouraged to refer to the references (\textit{and the references within}) below for more up-to-date and comprehensive discussions on Neutrino Physics, especially [Ref~\onlinecite{boyd, bouchandthesis, ma_arxiv, zuber2020neutrino, Dighe, King_2004, Mohapatra:2005wg}]. For a nice list of Open Questions in Neutrino Physics one may refer [Ref~\onlinecite{Smirnov_2004}]. 


\section*{References}
\bibliography{references}

\providecommand{\noopsort}[1]{}\providecommand{\singleletter}[1]{#1}%
\begin{thebibliography}{21}%
\makeatletter
\providecommand \@ifxundefined [1]{%
 \@ifx{#1\undefined}
}%
\providecommand \@ifnum [1]{%
 \ifnum #1\expandafter \@firstoftwo
 \else \expandafter \@secondoftwo
 \fi
}%
\providecommand \@ifx [1]{%
 \ifx #1\expandafter \@firstoftwo
 \else \expandafter \@secondoftwo
 \fi
}%
\providecommand \natexlab [1]{#1}%
\providecommand \enquote  [1]{``#1''}%
\providecommand \bibnamefont  [1]{#1}%
\providecommand \bibfnamefont [1]{#1}%
\providecommand \citenamefont [1]{#1}%
\providecommand \href@noop [0]{\@secondoftwo}%
\providecommand \href [0]{\begingroup \@sanitize@url \@href}%
\providecommand \@href[1]{\@@startlink{#1}\@@href}%
\providecommand \@@href[1]{\endgroup#1\@@endlink}%
\providecommand \@sanitize@url [0]{\catcode `\\12\catcode `\$12\catcode `\&12\catcode `\#12\catcode `\^12\catcode `\_12\catcode `\%12\relax}%
\providecommand \@@startlink[1]{}%
\providecommand \@@endlink[0]{}%
\providecommand \url  [0]{\begingroup\@sanitize@url \@url }%
\providecommand \@url [1]{\endgroup\@href {#1}{\urlprefix }}%
\providecommand \urlprefix  [0]{URL }%
\providecommand \Eprint [0]{\href }%
\providecommand \doibase [0]{https://doi.org/}%
\providecommand \selectlanguage [0]{\@gobble}%
\providecommand \bibinfo  [0]{\@secondoftwo}%
\providecommand \bibfield  [0]{\@secondoftwo}%
\providecommand \translation [1]{[#1]}%
\providecommand \BibitemOpen [0]{}%
\providecommand \bibitemStop [0]{}%
\providecommand \bibitemNoStop [0]{.\EOS\space}%
\providecommand \EOS [0]{\spacefactor3000\relax}%
\providecommand \BibitemShut  [1]{\csname bibitem#1\endcsname}%
\let\auto@bib@innerbib\@empty
\bibitem [{\citenamefont {Cowan~Jr}\ \emph {et~al.}(1956)\citenamefont {Cowan~Jr}, \citenamefont {Reines}, \citenamefont {Harrison}, \citenamefont {Kruse},\ and\ \citenamefont {McGuire}}]{cowan1956detection}%
  \BibitemOpen
  \bibfield  {author} {\bibinfo {author} {\bibfnamefont {C.~L.}\ \bibnamefont {Cowan~Jr}}, \bibinfo {author} {\bibfnamefont {F.}~\bibnamefont {Reines}}, \bibinfo {author} {\bibfnamefont {F.}~\bibnamefont {Harrison}}, \bibinfo {author} {\bibfnamefont {H.}~\bibnamefont {Kruse}},\ and\ \bibinfo {author} {\bibfnamefont {A.}~\bibnamefont {McGuire}},\ }\bibfield  {title} {\enquote {\bibinfo {title} {Detection of the free neutrino: a confirmation},}\ }\href {https://doi.org/10.1126/science.124.3212.103} {\bibfield  {journal} {\bibinfo  {journal} {Science}\ }\textbf {\bibinfo {volume} {124}},\ \bibinfo {pages} {103--104} (\bibinfo {year} {1956})}\BibitemShut {NoStop}%
\bibitem [{\citenamefont {Danby}\ \emph {et~al.}(1962)\citenamefont {Danby}, \citenamefont {Gaillard}, \citenamefont {Goulianos}, \citenamefont {Lederman}, \citenamefont {Mistry}, \citenamefont {Schwartz},\ and\ \citenamefont {Steinberger}}]{brookhaven}%
  \BibitemOpen
  \bibfield  {author} {\bibinfo {author} {\bibfnamefont {G.}~\bibnamefont {Danby}}, \bibinfo {author} {\bibfnamefont {J.-M.}\ \bibnamefont {Gaillard}}, \bibinfo {author} {\bibfnamefont {K.}~\bibnamefont {Goulianos}}, \bibinfo {author} {\bibfnamefont {L.~M.}\ \bibnamefont {Lederman}}, \bibinfo {author} {\bibfnamefont {N.}~\bibnamefont {Mistry}}, \bibinfo {author} {\bibfnamefont {M.}~\bibnamefont {Schwartz}},\ and\ \bibinfo {author} {\bibfnamefont {J.}~\bibnamefont {Steinberger}},\ }\bibfield  {title} {\enquote {\bibinfo {title} {Observation of high-energy neutrino reactions and the existence of two kinds of neutrinos},}\ }\href {https://doi.org/10.1103/PhysRevLett.9.36} {\bibfield  {journal} {\bibinfo  {journal} {Phys. Rev. Lett.}\ }\textbf {\bibinfo {volume} {9}},\ \bibinfo {pages} {36--44} (\bibinfo {year} {1962})}\BibitemShut {NoStop}%
\bibitem [{\citenamefont {et~al. [DONUT~Collaboration]}(2001)}]{KODAMA2001218}%
  \BibitemOpen
  \bibfield  {author} {\bibinfo {author} {\bibfnamefont {K.~K.}\ \bibnamefont {et~al. [DONUT~Collaboration]}},\ }\bibfield  {title} {\enquote {\bibinfo {title} {Observation of tau neutrino interactions},}\ }\href {https://doi.org/https://doi.org/10.1016/S0370-2693(01)00307-0} {\bibfield  {journal} {\bibinfo  {journal} {Physics Letters B}\ }\textbf {\bibinfo {volume} {504}},\ \bibinfo {pages} {218--224} (\bibinfo {year} {2001})}\BibitemShut {NoStop}%
\bibitem [{\citenamefont {de~Gouvêa}(2016)}]{Gouvêa_2016}%
  \BibitemOpen
  \bibfield  {author} {\bibinfo {author} {\bibfnamefont {A.}~\bibnamefont {de~Gouvêa}},\ }\bibfield  {title} {\enquote {\bibinfo {title} {Neutrino mass models},}\ }\href {https://doi.org/10.1146/annurev-nucl-102115-044600} {\bibfield  {journal} {\bibinfo  {journal} {Annual Review of Nuclear and Particle Science}\ }\textbf {\bibinfo {volume} {66}},\ \bibinfo {pages} {197–217} (\bibinfo {year} {2016})}\BibitemShut {NoStop}%
\bibitem [{\citenamefont {Zuber}(2020)}]{zuber2020neutrino}%
  \BibitemOpen
  \bibfield  {author} {\bibinfo {author} {\bibfnamefont {K.}~\bibnamefont {Zuber}},\ }\href {https://doi.org/10.1201/9781315195612} {\emph {\bibinfo {title} {Neutrino {Physics}}}},\ \bibinfo {edition} {3rd}\ ed.\ (\bibinfo  {publisher} {CRC Press},\ \bibinfo {year} {2020})\BibitemShut {NoStop}%
\bibitem [{\citenamefont {Dighe}(2020)}]{Dighe}%
  \BibitemOpen
  \bibfield  {author} {\bibinfo {author} {\bibfnamefont {A.}~\bibnamefont {Dighe}},\ }\bibfield  {title} {\enquote {\bibinfo {title} {Neutrino physics: An introduction},}\ }\href@noop {} {\  (\bibinfo {year} {2020})},\ \bibinfo {note} {lecture Notes, SERC school}\BibitemShut {NoStop}%
\bibitem [{\citenamefont {King}(2004)}]{King_2004}%
  \BibitemOpen
  \bibfield  {author} {\bibinfo {author} {\bibfnamefont {S.~F.}\ \bibnamefont {King}},\ }\bibfield  {title} {\enquote {\bibinfo {title} {Neutrino mass models},}\ }\href {https://doi.org/10.1088/0034-4885/67/2/R01} {\bibfield  {journal} {\bibinfo  {journal} {Reports on Progress in Physics}\ }\textbf {\bibinfo {volume} {67}},\ \bibinfo {pages} {107–157} (\bibinfo {year} {2004})},\ \bibinfo {note} {arXiv:hep-ph/0310204}\BibitemShut {NoStop}%
\bibitem [{\citenamefont {Aalseth}\ \emph {et~al.}(2004)\citenamefont {Aalseth} \emph {et~al.}}]{Aalseth:2004hb}%
  \BibitemOpen
  \bibfield  {author} {\bibinfo {author} {\bibfnamefont {C.}~\bibnamefont {Aalseth}} \emph {et~al.},\ }\bibfield  {title} {\enquote {\bibinfo {title} {{Neutrinoless double beta decay and direct searches for neutrino mass}},}\ }\href@noop {} {\  (\bibinfo {year} {2004})},\ \Eprint {https://arxiv.org/abs/hep-ph/0412300} {arXiv:hep-ph/0412300} \BibitemShut {NoStop}%
\bibitem [{\citenamefont {Ma}(2009)}]{ma_arxiv}%
  \BibitemOpen
  \bibfield  {author} {\bibinfo {author} {\bibfnamefont {E.}~\bibnamefont {Ma}},\ }\bibfield  {title} {\enquote {\bibinfo {title} {{Neutrino Mass: Mechanisms and Models}},}\ }\href@noop {} {\  (\bibinfo {year} {2009})},\ \Eprint {https://arxiv.org/abs/0905.0221} {arXiv:0905.0221 [hep-ph]} \BibitemShut {NoStop}%
\bibitem [{\citenamefont {Mohapatra}\ and\ \citenamefont {Senjanovi\ifmmode~\acute{c}\else \'{c}\fi{}}(1980)}]{MohapatraPhysRevLett}%
  \BibitemOpen
  \bibfield  {author} {\bibinfo {author} {\bibfnamefont {R.~N.}\ \bibnamefont {Mohapatra}}\ and\ \bibinfo {author} {\bibfnamefont {G.}~\bibnamefont {Senjanovi\ifmmode~\acute{c}\else \'{c}\fi{}}},\ }\bibfield  {title} {\enquote {\bibinfo {title} {Neutrino mass and spontaneous parity nonconservation},}\ }\href {https://doi.org/10.1103/PhysRevLett.44.912} {\bibfield  {journal} {\bibinfo  {journal} {Phys. Rev. Lett.}\ }\textbf {\bibinfo {volume} {44}},\ \bibinfo {pages} {912--915} (\bibinfo {year} {1980})}\BibitemShut {NoStop}%
\bibitem [{\citenamefont {Boyd}()}]{boyd}%
  \BibitemOpen
  \bibfield  {author} {\bibinfo {author} {\bibfnamefont {S.}~\bibnamefont {Boyd}},\ }\href {https://warwick.ac.uk/fac/sci/physics/staff/academic/boyd/stuff/neutrinolectures/} {\enquote {\bibinfo {title} {Neutrino lecture writeups},}\ }\bibinfo {note} {Course Notes}\BibitemShut {NoStop}%
\bibitem [{\citenamefont {Gelmini}\ and\ \citenamefont {Roncadelli}(1981)}]{type3seesaw}%
  \BibitemOpen
  \bibfield  {author} {\bibinfo {author} {\bibfnamefont {G.}~\bibnamefont {Gelmini}}\ and\ \bibinfo {author} {\bibfnamefont {M.}~\bibnamefont {Roncadelli}},\ }\bibfield  {title} {\enquote {\bibinfo {title} {Left-handed neutrino mass scale and spontaneously broken lepton number},}\ }\href {https://doi.org/https://doi.org/10.1016/0370-2693(81)90559-1} {\bibfield  {journal} {\bibinfo  {journal} {Physics Letters B}\ }\textbf {\bibinfo {volume} {99}},\ \bibinfo {pages} {411--415} (\bibinfo {year} {1981})}\BibitemShut {NoStop}%
\bibitem [{Note1()}]{Note1}%
  \BibitemOpen
  \bibinfo {note} {Charge conservation rules out the possibility for a singlet.}\BibitemShut {Stop}%
\bibitem [{\citenamefont {Bouchand}(2012)}]{bouchandthesis}%
  \BibitemOpen
  \bibfield  {author} {\bibinfo {author} {\bibfnamefont {R.}~\bibnamefont {Bouchand}},\ }\href {http://kth.diva-portal.org/smash/get/diva2:517225/FULLTEXT01} {\enquote {\bibinfo {title} {Radiative neutrino mass generation and renormalization group running in the ma-model},}\ } (\bibinfo {year} {2012}),\ \bibinfo {note} {master Thesis}\BibitemShut {NoStop}%
\bibitem [{\citenamefont {Zee}(1980)}]{zee_wolfstein}%
  \BibitemOpen
  \bibfield  {author} {\bibinfo {author} {\bibfnamefont {A.}~\bibnamefont {Zee}},\ }\bibfield  {title} {\enquote {\bibinfo {title} {{A Theory of Lepton Number Violation, Neutrino Majorana Mass, and Oscillation}},}\ }\href {https://doi.org/10.1016/0370-2693(80)90349-4} {\bibfield  {journal} {\bibinfo  {journal} {Phys. Lett. B}\ }\textbf {\bibinfo {volume} {93}},\ \bibinfo {pages} {389} (\bibinfo {year} {1980})},\ \bibinfo {note} {[Erratum: Phys.Lett.B 95, 461 (1980)]}\BibitemShut {NoStop}%
\bibitem [{\citenamefont {He}(2004)}]{He:2003ih}%
  \BibitemOpen
  \bibfield  {author} {\bibinfo {author} {\bibfnamefont {X.-G.}\ \bibnamefont {He}},\ }\bibfield  {title} {\enquote {\bibinfo {title} {{Is the Zee model neutrino mass matrix ruled out?}}}\ }\href {https://doi.org/10.1140/epjc/s2004-01669-8} {\bibfield  {journal} {\bibinfo  {journal} {Eur. Phys. J. C}\ }\textbf {\bibinfo {volume} {34}},\ \bibinfo {pages} {371--376} (\bibinfo {year} {2004})},\ \Eprint {https://arxiv.org/abs/hep-ph/0307172} {arXiv:hep-ph/0307172} \BibitemShut {NoStop}%
\bibitem [{\citenamefont {Zee}(1986)}]{Zee:1985id}%
  \BibitemOpen
  \bibfield  {author} {\bibinfo {author} {\bibfnamefont {A.}~\bibnamefont {Zee}},\ }\bibfield  {title} {\enquote {\bibinfo {title} {{Quantum Numbers of Majorana Neutrino Masses}},}\ }\href {https://doi.org/10.1016/0550-3213(86)90475-X} {\bibfield  {journal} {\bibinfo  {journal} {Nucl. Phys. B}\ }\textbf {\bibinfo {volume} {264}},\ \bibinfo {pages} {99--110} (\bibinfo {year} {1986})}\BibitemShut {NoStop}%
\bibitem [{\citenamefont {Babu}(1988)}]{Babu:1988ki}%
  \BibitemOpen
  \bibfield  {author} {\bibinfo {author} {\bibfnamefont {K.~S.}\ \bibnamefont {Babu}},\ }\bibfield  {title} {\enquote {\bibinfo {title} {{Model of 'Calculable' Majorana Neutrino Masses}},}\ }\href {https://doi.org/10.1016/0370-2693(88)91584-5} {\bibfield  {journal} {\bibinfo  {journal} {Phys. Lett. B}\ }\textbf {\bibinfo {volume} {203}},\ \bibinfo {pages} {132--136} (\bibinfo {year} {1988})}\BibitemShut {NoStop}%
\bibitem [{\citenamefont {Babu}\ and\ \citenamefont {Ma}(1988)}]{babu_ma_phyrev}%
  \BibitemOpen
  \bibfield  {author} {\bibinfo {author} {\bibfnamefont {K.~S.}\ \bibnamefont {Babu}}\ and\ \bibinfo {author} {\bibfnamefont {E.}~\bibnamefont {Ma}},\ }\bibfield  {title} {\enquote {\bibinfo {title} {Natural hierarchy of radiatively induced majorana neutrino masses},}\ }\href {https://doi.org/10.1103/PhysRevLett.61.674} {\bibfield  {journal} {\bibinfo  {journal} {Phys. Rev. Lett.}\ }\textbf {\bibinfo {volume} {61}},\ \bibinfo {pages} {674--677} (\bibinfo {year} {1988})}\BibitemShut {NoStop}%
\bibitem [{\citenamefont {Mohapatra}\ \emph {et~al.}(2007)\citenamefont {Mohapatra} \emph {et~al.}}]{Mohapatra:2005wg}%
  \BibitemOpen
  \bibfield  {author} {\bibinfo {author} {\bibfnamefont {R.~N.}\ \bibnamefont {Mohapatra}} \emph {et~al.},\ }\bibfield  {title} {\enquote {\bibinfo {title} {{Theory of neutrinos: A White paper}},}\ }\href {https://doi.org/10.1088/0034-4885/70/11/R02} {\bibfield  {journal} {\bibinfo  {journal} {Rept. Prog. Phys.}\ }\textbf {\bibinfo {volume} {70}},\ \bibinfo {pages} {1757--1867} (\bibinfo {year} {2007})},\ \Eprint {https://arxiv.org/abs/hep-ph/0510213} {arXiv:hep-ph/0510213} \BibitemShut {NoStop}%
\bibitem [{\citenamefont {Smirnov}(2004)}]{Smirnov_2004}%
  \BibitemOpen
  \bibfield  {author} {\bibinfo {author} {\bibfnamefont {A.~Y.}\ \bibnamefont {Smirnov}},\ }\bibfield  {title} {\enquote {\bibinfo {title} {Neutrino physics: open theoretical questions},}\ }\href {https://doi.org/10.1142/S0217751X0401910X} {\bibfield  {journal} {\bibinfo  {journal} {International Journal of Modern Physics A}\ }\textbf {\bibinfo {volume} {19}},\ \bibinfo {pages} {1180–1197} (\bibinfo {year} {2004})}\BibitemShut {NoStop}%
\end{thebibliography}%
\onecolumngrid
\addcontentsline{toc}{section}{Appendix}
\appendix
\section{\label{sec:charge_conjugation}Charge Conjugation}
``Charge conjugation operator" flips the signs of all the charges (or quantum numbers). It changes a particle into an anti-particle, and vice versa. The relevant operator is:
\begin{equation}
    \hat{C} \ket{\Psi} = C \ket{\overline {\Psi}}
\end{equation}
where $C$ on the right hand side of the equation is the charge conjugation eigenvalue. One can show that the form for the charge conjugation operator, using Dirac gamma matrices, is
\begin{equation}
    \hat{C} = i \gamma ^2 \gamma ^0
\end{equation}
If \(\psi\) is the spinor field of a free neutrino, then the charge-conjugated field \(\psi^c\) can be shown to be
\begin{equation}
    \psi^c = C \psi ^*
\end{equation}
\(\psi ^*\) is charge conjugated field. \cite{boyd}
\section{\label{sec:helicity}Helicity}
The fact that we can find spin eigenvalues for states in which the particles are travelling along the spin-direction indicates that the quantity we need is not \textit{spin} but \textit{helicity}. The \textsc{helicity is defined as the projection of the spin along the direction of motion}:
\begin{equation}
    \hat{h} = \bm{\Sigma} \cdot \hat{\bm{p}}  = 2 \bm{S}\cdot\hat{\bm{p}} = \begin{pmatrix}
        \sigma & 0\\
        0 & \sigma
    \end{pmatrix} \cdot \hat{\bm{p}}
\end{equation}
and has eigenvalues equal to +1 (called \textit{right-handed} where the spin vector is aligned in the same
direction as the momentum vector) or -1 (called \textit{left-handed} where the spin vector is aligned in the opposite direction as the momentum vector)

It can be shown that the helicity \textit{does} commute with the Hamiltonian and so one can find eigenstates that are simultaneously states of helicity and the Hamiltonian. The problem, and it is a big problem, is that helicity is not Lorentz invariant in the case of a massive particle. If the particle is massive it is possible to find an inertial reference frame in which the particle is going in the opposite direction. This does not change the direction of the spin vector, so the helicity can change sign.

\textsc{The helicity is Lorentz invariant only in the case of massless particles.}
\section{\label{sec:chirality}Chirality}
It would be good if we can find a operator that is Lorentz invariant, rather than commuting with the Hamiltonian. In general wave functions in the Standard Model are eigenstates of a Lorentz invariant quantity called the \textit{chirality}. The chirality operator is \(\gamma ^5\) and it does not commute with the Hamiltonian. 
The chirality is identical to the helicity in the limit that \(E \gg m\), or that the particle is massless. For a massive particle, this is no longer true.

In general the eigenstates of the chirality operator are
\begin{equation}
    \begin{gathered}
        \gamma ^5 u_R = + u_R, ~ \gamma ^5 u_L = - u_L\\
        \gamma ^5 v_R = - v_R, ~ \gamma ^5 v_L = + v_L
    \end{gathered}
\end{equation}
where we define \(u_R\) and \( u_R\) are right and left-handed chiral states. One can define the projection operators as 
\begin{equation}
    \begin{gathered}
    P_L = \frac{1}{2} (1 - \gamma^5), \quad P_R = \frac{1}{2} (1 + \gamma^5)
\end{gathered}
\end{equation}
uch that $P_L$ projects outs the left-handed chiral particle states and right-handed chiral anti-particle states. $P_R$ projects out the right-handed chiral particle states and left-handed chiral anti-particle states.

Any spinor can be written in terms of it’s left- and right-handed chiral states:
\begin{equation}
    \psi = (P_R + P_L)\psi  = P_R \psi + P_L\psi = \psi _R + \psi _L
\end{equation}
using the properties of projection operators, it's easy to show that for some fields \(\psi\) and \(\psi\)
\begin{equation}
    \overline {\psi}_L \gamma_\mu \phi _R = \overline {\psi}_R \gamma_\mu \phi _L = 0
\end{equation}
So left-handed chiral particles couple only to left-handed chiral fields, and right-handed chiral fields couple to right-handed chiral fields.

One must be very careful with how one interprets this statement. What it does not say is that
there are left-handed chiral electrons and right-handed chiral electrons which are distinct particles. The ``particles" are those states which propagate with fixed mass under the Dirac equation. What this means is that the Standard Model does not mix chiralities or, to put it another way, no fundamental interaction is capable of turning a left-chiral particle into a right-chiral particle. Useful though it is when describing the interaction of fields in the Standard Model, chirality is not conserved in the propagation of a free particle. In fact the chiral states \(\phi _L\) and \(\phi_R\) do not even satisfy the Dirac equation. Since chirality is not a good quantum number it can evolve with time. A massive particle starting off as a completely left-handed chiral state can evolve a right-handed chiral component. By contrast, helicity is a conserved quantity during free particle propagation. Only in the case of massless particles, for which helicity and chirality are identical and are conserved in free-particle propagation, can left- and right-handed particles be considered distinct. For neutrinos this mostly holds.

\end{document}